\documentclass[twocolumn,showpacs,superscriptaddress]{revtex4-2}
\usepackage{amsmath,amssymb, graphicx, subcaption, hyperref, braket, float}
\usepackage{blindtext}

\counterwithout{figure}{section}
\usepackage[export]{adjustbox}
\usepackage{natbib}
\graphicspath{{figure/}}
\usepackage[T1]{fontenc}
\usepackage{ae,aecompl}
\usepackage{comment}
\usepackage{xcolor}

\begin{document}
	\title{ 
Stabilization of quantum properties under intrinsic decoherence in presence of external magnetic fields}
\author{Essalha Chaouki}
\email{essalha.chaouki-etu@etu.univh2c.ma}
\email{1762100@uab.cat}

\affiliation{Laboratory of High Energy Physics and Condensed Matter, Department of Physics, Faculty of Sciences Ain Chock, Hassan II University, P.O. Box 5366 Maarif, Casablanca 20100,  Morocco}
\affiliation{F\'{\i}sica Te\`{o}rica: Informaci\'{o} i Fen\`{o}mens Qu\`{a}ntics, %
	Departament de F\'{\i}sica, Universitat Aut\`{o}noma de Barcelona, 08193 Bellaterra, Spain}
 
	\author{Anna Sanpera}
    \email{Anna.Sanpera@uab.cat}
 \affiliation{F\'{\i}sica Te\`{o}rica: Informaci\'{o} i Fen\`{o}mens Qu\`{a}ntics, %
	Departament de F\'{\i}sica, Universitat Aut\`{o}noma de Barcelona, 08193 Bellaterra, Spain}
\affiliation{ICREA, Pg. Llu\'is Companys 23, 08010 Barcelona, Spain}

	\author{Mostafa Mansour}
	\email{mostafa.mansour.fpb@gmail.com}
	\affiliation{Laboratory of High Energy Physics and Condensed Matter, Department of Physics, Faculty of Sciences Ain Chock, Hassan II University, P.O. Box 5366 Maarif, Casablanca 20100,  Morocco}

	\begin{abstract}
		The dynamical behavior of quantum state properties under intrinsic decoherence models can be modified by the presence of external magnetic fields. Although generically external magnetic fields are detrimental to preserve quantumness in the presence of intrinsic decoherence, judicious adjustment of the magnetic field can stabilize such features. This stabilization arises from novel resonances between energy eigenstates resulting from the presence of an external magnetic field. Here, we present our findings using as a model system two spin 1-particles confined in a double-well potential under intrinsic decoherence. We stress, however, that our results are generic and independent on the used model.
	\end{abstract}
	\maketitle
	\section{Introduction} 

Quantum resource theories have emerged in the last years as a unified framework to characterize and quantified the potential use of diverse intrinsic quantum properties\cite{chitambar2019quantum,streltsov2017colloquium,guhne2002detection}. Within this framework, quantum coherence and entanglement, both pivotal hallmarks characterizing nonclassical systems, have been thoroughly analyzed due to their crucial role for a spectrum of applications in quantum information processing\cite{bennett2000quantum,bennett1996concentrating,amico2008entanglement,sen2006quantum}, extending its influence into burgeoning fields like quantum metrology \cite{giovannetti2004quantum}, quantum thermodynamics \cite{santos2019role}, quantum biology \cite{li2012witnessing} and generic quantum technologies.
As a result, there has been extensive research that delves into the complex interplay among the different parameters that influence entanglement and coherence. For example, the interrelation between coherence and entanglement has been widely analyzed \cite{streltsov2015measuring, ma2016converting}, but also the relation between entanglement, entropy and mixedness \cite{wei2003maximal, bandyopadhyay2012entanglement}, or coherence and mixedness \cite{cheng2015complementarity, singh2015maximally} has become a crucial focus of the present research. 

Coherence, entanglement, and purity are, however, highly susceptible to the effects of decoherence and dissipation\cite{obada2002influence,obada2007entropy}. Notably, even in closed quantum systems, the effects of decoherence can be included using the so-called \textit{intrisinc decoherence} (ID) models. In \cite{milburn1991intrinsic}, Milburn introduced a key model of ID by altering the Schrödinger equation at sufficient small time steps. The effects of such a model in various quantum systems have been largely explored \cite{chaouki2022dynamics,he2006influence, wei2009effects,oum1,mendoncca2017maximally,liang2008effect}. Here, we aim at understanding how ID affects entanglement, coherence, and purity for a two-qutrit system under the effect of external magnetic fields \cite{jaghouri2018thermal}. On the one hand, such a model mimics the minimum instance of a spin-1 ultracold atoms in optical lattices\cite{lewenstein2012ultracold, Dechiara} under the action of external magnetic fields. 
On the other hand, qutrits have demonstrated superior efficacy over qubits in several tasks. For instance, there is an advantage in using qutrits for quantum error correction fault-tolerant schemes, enabling the creation of compact and more efficient codes that lack direct equivalents in qubit systems \cite{qut1}. Also, quantum cryptography employing 3-dimensional quantum states for optimal eavesdropping has been identified as more resilient against symmetric attacks as compared to protocols utilizing 2-dimensional states \cite{qut2}. In the same instance, external magnetic fields are ubiquitous and usually detrimental, causing faster decoherence. Here, however, we show they can control system properties by resonance occurring between energy eigenstates, thus preserving the quantum features of the system.

The structure of the paper is as follows. In Section \ref{sec2}, we introduce the model under examination, the simplified spin-1 Bose-Hubbard model under intrinsic decoherence.  Such a model allows one to obtain a closed expression of the time-dependent density operator. In Section \ref{sec3}, we review the quantities utilized in this study to analyze the time dependence of coherence, entanglement, purity, and entropy.  Section \ref{sec4}, comprehensively explores the behaviors exhibited by these quantifiers and highlights the role that resonances play. Finally, in Section \ref{sec5} we summarize our results and present some open questions.
\section{ The Bose-Hubbard Hamiltonian under intrinsic decoherence \label{sec2}}
A system of bosons trapped in a deep optical lattice potential is well described by a Bose-Hubbard model, consisting of two competing terms: the hopping between lattice sites and the
repulsive interaction produced by local scattering. Bosonic interactions, due to two-body s-wave collisions, are sensitive to the spin degree of freedom. For spin-1 particles,
two possible scattering channels open; one with total spin $s = 0$, 
and another with total spin $s = 2$, with corresponding
scattering lengths $a_0$ and $a_2$.  
In the so-called Mott regime, where each well accommodates only one atom (or an integer number of them), the system is described by the so-called generalized quadratic Heisenberg or bilinear-biquadratic spin-1 model (see e.g.,\cite{lewenstein2012ultracold}), where tunneling is included up to second order. In presence of external magnetic fields, the Hamiltonian reads
\begin{eqnarray} 
\label{Hamiltonian}
H &=& \chi + \sum_i \left[J \left(\vec{S}_i \cdot \vec{S}_{i+1}\right) + {K} 
\left( \vec{S}_i \cdot \vec{S}_{i+1}\right)^2 \right] \\ \nonumber
&+& \sum_i B_z {S}_{i z} 
\end{eqnarray}
Here, $\chi=J-K$, $K=-\frac{2 {{\texttt{t}}}^2}{3 U_2}-\frac{4 {\texttt t}^2}{U_0}$, and $J=-\frac{2  {\texttt t}^2}{U_2}$, where the hopping term $\texttt t$ is assumed to be spin-independent. The Hubbard repulsion potentials, $U_s (s=0,2)$, are associated with the eigenvalues of the total spin $\vec{S}$ and depend on the corresponding scattering lengths. Since $\chi$ is constant, it can be disregarded. Thus, $J$ represents the Heisenberg interaction strength (linear coupling), and the parameter $K$ denotes the nonlinear coupling constant, also known as the biquadratic term. The external magnetic field along the $z$-axis is represented by $B_z$.

From now on, we focus on the two qutrit-system, considering only two spin-1 atoms in a double well. In this configuration, the eigenvectors $\ket{E_i}$ and eigenvalues $E_i$ of the Hamiltonian  read:
\begin{align*}
\ket{E_1} &=\frac{1}{\sqrt{3}}(|2,0\rangle-|1,1\rangle+|0,2\rangle) 
&\rightarrow E_1 &=4 {K}-2{J}\\
\ket{E_2} &=\frac{1}{\sqrt{2}}(|2,0\rangle-|0,2\rangle) &\rightarrow E_2&= {K}-{J}\\
\ket{E_3}& =\frac{1}{\sqrt{2}}(|1,2\rangle-|2,1\rangle) &\rightarrow E_3&= K-J-B_z\\
\ket{E_4} &=\frac{1}{\sqrt{2}}(|0,1\rangle-|1,0\rangle)&\rightarrow E_4 &=K-J+B_z\\
\ket{E_5} &=\frac{1}{\sqrt{6}}(|2,0\rangle+2|1,1\rangle+|0,2\rangle)&\rightarrow E_5&= K+J\\
\ket{E_6} &=|2,2\rangle&\rightarrow E_6&= K+J-2B_z\\
\ket{E_7} &=\frac{1}{\sqrt{2}}(|1,2\rangle+|2,1\rangle)&\rightarrow E_7&=K+J-B_z\\
\ket{E_8} &=\frac{1}{\sqrt{2}}(|0,1\rangle+|1,0\rangle)&\rightarrow  E_8&= K+J+B_z\\
\ket{E_9} &=|0,0\rangle&\rightarrow E_9&=K+J+2B_z
\label{eigenstates}
\end{align*}

We can now include decoherence in our system by using the ID model of \cite{milburn1991intrinsic}. Milburn proposed a beautiful and simplified model of decoherence that does not require a reservoir or conventional energy dissipation commonly associated with decay. In this approach, on a sufficiently small time scale, the quantum system experiences a series of successive unitary transformations rather than a single unitary evolution. The inverse of such a time scale corresponds to the intrinsic decoherence rate. The Schrodinger equation is recovered to zeroth order in the expansion parameter; however, higher-order corrections lead to a loss of coherence in the energy basis. Specifically, the dynamical equation reduces to:
\begin{equation}
\frac{d\varrho(t)}{dt} = \frac{1}{\gamma}\Bigl[e^{-i\gamma H}\varrho(t) e^{i\gamma H}-\varrho(t)\Bigr].
\label{milburn}
\end{equation}

The above equation describes the evolution of state $\varrho(t)$ with respect to the Hamiltonian $H$, under intrinsic decoherence given by the decoherence rate $\gamma$.  
In the limit where $\gamma^{-1}$ approaches infinity, the equation simplifies to the von Neumann equation, which characterizes the evolution of an isolated quantum system as $(\frac{d\varrho(t)}{dt}=-i [H,\varrho(t)])$, while for $\gamma$ finite, and developing up to second order, one arrives to the following expression:
\begin{equation}
\frac{d\varrho(t)}{dt}=-i [H,\varrho(t)]-\frac{\gamma}{2}[H,[H,\varrho(t)]].
\label{milburn1}
\end{equation}
The appearance of the term $\frac{\gamma}{2}[H,[H,\varrho(t)]]$ indicates the presence of non-unitary dynamics caused by ID.
The solution to equation \eqref{milburn1} can be derived employing the Kraus operators formalism, $\texttt{{\cal M}}_p$,\cite{milburn1991intrinsic}
\begin{equation*}
	\varrho(t)= \sum^{\infty}_{p=0} \texttt{{\cal M}}_{p}(t)
  \varrho(0) \texttt{{\cal M}}_{p}^{\dagger}(t),
\end{equation*}
wherein the term $\varrho{(0)}$ denotes the density matrix corresponding to $t=0$, and 
$\texttt{{\cal M}}_p(t)$,  
the time-dependent Kraus operators: 
\begin{equation}
	\texttt{\cal M}_p(t) = 
    \sqrt{\frac{(\gamma t)^p}{p!}} {H}^p\exp\left(-i{H}t\right) \exp\left(-\frac{\gamma t}{2}{H}^{2}\right),
 \end{equation}
where $\sum^{\infty}_{p=0} \texttt{{\cal M}}^{\dagger}_p(t) \texttt{{\cal M}}_p(t)  =\mathbb{I}$. Using the above expressions, one arrives at a compact formula for the time evolution equation of the density matrix under Hamiltonian dynamics:
\begin{eqnarray}
\displaystyle
\varrho(t) &=& \sum_{j,k} \exp\left(-\frac{\gamma t}{2}(E_k-E_j)^{2}- it(E_k-E_j)\right) \nonumber\\ &\times& \bra{\mathbf{E}_k}\varrho{(0)}\ket{\mathbf{E}_j} \ket{\mathbf{E}_k}\bra{\mathbf{E}_j}.
\label{sol}
\end{eqnarray}
Equation \eqref{sol} describes how quantum coherence is degraded as the system evolves in the energy eigenbasis of the Hamiltonian. Thus, the steady-state solution of the system depends on the projection of the initial state $\varrho{(0)}$, onto the eigenstates of the Hamiltonian. At sufficient large times, intrinsic decoherence effects, modeled by $\exp\left(-\frac{\gamma t}{2}(E_k-E_j)^{2}- it(E_k-E_j)\right)$, typically suppress off-diagonal terms over time by introducing decay. However, under specific conditions, as we shall demonstrate later (appendix\ref{appendix:energy_basis}), survival of these terms can be guaranteed, preserving the system's quantum properties. 

For convenience, in what follows, we choose the initial state to be the so-called isotropic state. 
\begin{eqnarray}
\varrho_{\text{iso}}(0)=\frac{(1-p)}{9} \mathbb{I}_3 \otimes \mathbb{I}_3+p|\psi\rangle\langle\psi|,
\label{state0} 
\end{eqnarray}
where $|\psi\rangle=(|0,0\rangle+|1,1\rangle+|2,2\rangle) / \sqrt{3}$ is the maximally entangled state, $p \in[0,1]$ and $\mathbb{I}_n$ denotes the identity operator in $n$-dimensional Hilbert space.
These states, also known as qutrit Werner states,
have attracted considerable attention from various perspectives (see \cite{tsokeng2018free} and references therein). Moreover, they allow for a straight characterization of their purity and entanglement and are highly symmetric, making the characterization of quantum features more straightforward. However, we stress that our analysis is generic and does not depend on the initial chosen state, though the details are, of course, dependent on it. 

Despite the entanglement characterization of generic two-qutrit states is unknown, this is not the case for a Werner state. There exist a critical threshold, $p_c=1/4 $, discriminating between the separable states 
$p\leq p_c$, and the entangled ones $p>1/4$ \cite{caves2000qutrit}.
After performing a straightforward calculation (\ref{sol}), the density matrix written in the computational basis is highly symmetric, decoupling a time-independent subspace while preserving the dynamical evolution into its complementary subspace, as indicated below: 
\begin{equation}         
\label{Sigma}
\varrho(t) =\begin{pmatrix}
\varrho_{1,1} & 0 & \varrho_{1,3} & 0 & \varrho_{1,5} & 0 & \varrho_{1,7} & 0 & \varrho_{1,9} \\
 0 & \varrho_{2,2} & 0 & 0 & 0 & 0 & 0 & 0 & 0 \\
\varrho^{*}_{1,3} & 0 & \varrho_{3,3} & 0 & \varrho_{3,5} & 0 & \varrho_{3,7} & 0 & \varrho_{3,9} \\
 0 & 0 & 0 & \varrho_{4,4} & 0 & 0 & 0 & 0 & 0 \\
 \varrho^{*}_{1,5} & 0 & \varrho^{*}_{3,5} & 0 & \varrho_{5,5} & 0 & \varrho_{5,7} & 0 & \varrho_{5,9} \\
 0 & 0 & 0 & 0 & 0 & \varrho_{6,6} & 0 & 0 & 0 \\
\varrho^{*}_{1,7} & 0 & \varrho^{*}_{3,7} & 0 & \varrho^{*}_{5,7} & 0 & \varrho_{7,7} & 0 & \varrho_{7,9} \\
 0 & 0 & 0 & 0 & 0 & 0 & 0 & \varrho_{8,8} & 0 \\
 \varrho^{*}_{1,9} & 0 & \varrho^{*}_{3,9} & 0 & \varrho^{*}_{5,9} & 0 & \varrho^{*}_{7,9} & 0 & \varrho_{9,9}\nonumber
\end{pmatrix},
\end{equation}
where
\begin{equation}
\begin{split}
\varrho_{1,1}&=\varrho_{9,9}=\frac{2 p+1}{9};\\
\varrho_{1,5}&=\frac{p}{9} (2 \texttt{{\cal a}} +\texttt{{\cal b}});\\
\varrho_{3,9}&=\varrho_{7,9}=\frac{p }{9} (\texttt{\cal a} -\texttt{\cal c});\\
\end{split}
\quad
  \begin{split}
\varrho_{1,3}&=\varrho_{1,7}=\frac{p}{9}  (\texttt{\cal a} -\texttt{\cal b});\\
\varrho_{1,9}&=\frac{p}{3}  e^{-4 {B_z} t (2 {B_z} \gamma +i)};\\
\end{split}
\end{equation}

\begin{subequations}\label{elements}
\begin{align*}
\displaystyle
\varrho_{2,2}&=\varrho_{4,4}=\varrho_{6,6}=\varrho_{8,8}=\frac{1-p}{9};\\
\varrho_{3,3}&=\varrho_{7,7}=\frac{\texttt{{\cal d}}}{27} (-\epsilon  p-(p-3) \texttt{{\cal d}} -p);\\
\varrho_{3,5}&=\varrho^{*}_{5,7}=\frac{p \texttt{{\cal d}}}{27}   (-2 \epsilon +\texttt{{\cal d}} +1);\\
\varrho_{3,7}&=\varrho^{*}_{3,7}=-\frac{p \texttt{{\cal d}}}{27} (\epsilon -2 \texttt{{\cal d}} +1);\\
\varrho_{5,5}&=\frac{\texttt{{\cal d}}}{27}   (2 \epsilon  p+(2 p+3) \texttt{{\cal d}} +2 p);\\
\varrho_{5,9}&=\frac{p}{9} (2 \texttt{{\cal a}} +\texttt{{\cal c}}),\\
\end{align*}
\end{subequations}
with
\begin{align*}
\texttt{{\cal a}}&=\exp \left(-2 {B_z} t ({B_z} \gamma +i)\right), \quad \epsilon=\exp \left(6 i t ({J}-{K})\right), \\
\texttt{{\cal b}}&=\exp \left(- t (2 {B_z}+3({J}-{K})) (\frac{\gamma}{2}(2 {B_z}+3({J}-{K}))+ i)\right),\\
\texttt{{\cal c}}&=\exp \left(- t (2 {B_z}-3({J}-{K})) (\frac{\gamma}{2}(2 {B_z}-3({J}-{K}))+ i)\right),\\
\texttt{{\cal d}}&=\exp \left(\frac{9}{2}\gamma t ({J}-{K})^2+ 3 it({J}-{K})\right).
\end{align*}
\section{Quantifying Coherence, Mixedness, and Entanglement \label{sec3}}
\begin{figure}[H]
\centering
\includegraphics[width=0.4\textwidth]{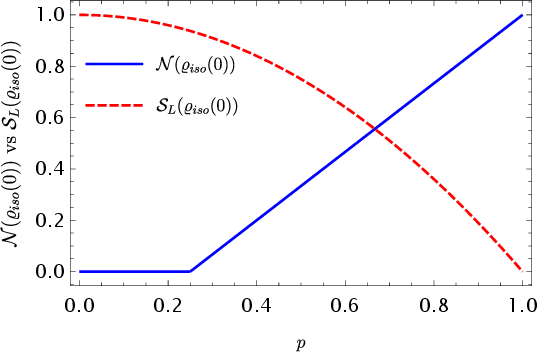}
    \caption{Linear Entropy and Negativity of the isotropic state, $\varrho_{iso}$, as a function of the purity $p$.}
    \label{fig:the Qutrit-Qutrit Isotropic}
\end{figure}

Let us briefly review here the metrics used to characterize entanglement, coherence, and linear entropy within the two-qutrit system. \\
\noindent \textbf{$l_1$-norm of coherence:}
Here, we use the $l_1$ coherence norm to quantify quantum coherence in a two-qutrit state, which has good properties as shown in \cite{baumgratz2014quantifying}, defined as follows:
\begin{equation}
\mathcal{C}_{l_{1}}(\varrho) = \min_{\xi} |\varrho - \xi|
\end{equation}
This $l_1$-norm coherence can be quantified by the off-diagonal elements of the density matrix:
\begin{equation}
\mathcal{C}_{l_{1}}(\varrho) = \sum_{i\neq j}|\varrho_{i,j}|
\end{equation}
Here, $|\varrho_{i,j}|$ represents the absolute value of the matrix elements of $\varrho$. Obviously, coherence measures depend on the choice of basis. Here, we restrict ourselves to the computational basis. 
Returning to the time-dependent density matrix of the qutrit-qutrit system under intrinsic decoherence, as defined in Equation \eqref{Sigma}, we can determine the $l_{1}$-norm coherence by summing the off-diagonal elements of this matrix. This expression can be formulated as follows:
\begin{eqnarray}
\displaystyle
C_{l_{1}}(\varrho(t))&=&\dfrac{1}{27}p \left( \right. 6|\texttt{{\cal a}}^*-\texttt{{\cal b}}^*|+3|2 \texttt{{\cal a}}^*+\texttt{{\cal b}}^*|+6|\texttt{{\cal a}}-\texttt{{\cal b}}|\nonumber
\\
&+&3| 2 \texttt{{\cal a}}+\texttt{{\cal b}}|+6|\texttt{{\cal a}}^*-\texttt{{\cal c}}^*|+3|2 \texttt{{\cal a}}^*+\texttt{{\cal c}}^*|\nonumber
\\
&+&6|\texttt{{\cal a}}-\texttt{{\cal c}}|+3|2 \texttt{{\cal a}}+\texttt{{\cal c}}|+18 e^{-8 {B_z}^2 \gamma t}\nonumber
\\
&+&2 e^{-\frac{9}{2} \gamma  t ({J}-{K})^2}(|1+\epsilon-2 \texttt{{\cal d}}|+|1-2 \epsilon+\texttt{{\cal d}}|\nonumber
\\
&+&| -2+\epsilon+\texttt{{\cal d}}|)
\left. \right).
\label{x4}
\end{eqnarray}
\\
\noindent\textbf{Negativity:} Negativity is frequently used to quantify bipartite entanglement \cite{vidal2002computable,eisert1999comparison,plenio2007introduction}, regardless of the dimensionality of the Hilbert spaces involved. 
\begin{equation}        \label{N}
N\left(\varrho_{A B}\right)=\frac{\left\|\varrho_{A B}^{\Gamma}\right\|-1}{2},
\end{equation}
where $\varrho_{A B}^{\Gamma}$ refers to the partial transpose of $\varrho_{A B}$ (irrespectively of which subsystem)  where the term $\|X\|_1=\operatorname{Tr}\left(\sqrt{X^{\dagger} X}\right)$ is the trace norm of an Hermitian operator $X$.
It is important to note that in the case of maximally entangled states, the negativity is precisely half $(N(\rho)=1 / 2)$, while completely separable (factorizable) states are characterized by $N(\rho)=0$. For convenience, we rescale by a factor 2, the negativity definition, denoted as $\mathcal{N}(\rho)=2 \times N(\rho)$. The negativity $\mathcal{N}\left(\varrho(\texttt{{t}})\right)$ linked to the time-varying state \eqref{Sigma} as:
\begin{equation} \label{x1}
\mathcal{N}\left(\varrho(\texttt{({t}})\right)=\sum_{\substack{i=1}}^{9}\left[\max\{0,-\hat{\mu}_i(t)\}\right].
\end{equation}
In this equation, $\hat{\mu_i}(t)$ denotes the eigenvalues of $\varrho_{A B}^{T_A}(t)$. Furthermore, the negativity of the initial state $\varrho_{\text{iso}}(0)$ \eqref{state0} can be directly obtained from \eqref{N} and can be succinctly expressed as follows:
\begin{eqnarray}
\displaystyle
\label{n_l0}
\mathcal{N}( \varrho_{\text{iso}}(0))=3(\max\{0,\frac{1}{9} (4 p-1)\}).
\end{eqnarray} 
\\
\noindent\textbf{Linear entropy:}
 The linear entropy denoted by $\mathcal{S}_L(\varrho)$\cite{wei2003maximal, li2006thermal}, is defined as 
\begin{center} 
\begin{equation}   \label{6}
\mathcal{S}_L=\frac{N}{N-1}\left[1-\operatorname{Tr}\left(\varrho^2\right)\right],
\end{equation}
\end{center}
where $N$ is the dimension of the system, i.e., dim ($\mathcal H_{AB})$. Clearly, linear entropy is a measure of the purity of the state.
Now, when considering our qutrit-qutrit system described by $\varrho(t)$ in Eq. \eqref{Sigma}  
the linear entropy corresponding to the time-dependent state \eqref{Sigma} can be expressed as follows:
\begin{eqnarray}
\displaystyle
\label{x2}
\mathcal{S}_L(\varrho(t))&=&\frac{1}{36} p^2 \left( \right.-9 e^{-16 {B_z}^2 \gamma t}-12 e^{-4 {B_z}^2 \gamma t}\nonumber
\\
&-&3 e^{- \gamma t (2 {B_z}+3 ({J}-{K}))^2} - 3 e^{- \gamma t (2 {B_z}-3 ({J}-{K}))^2}\nonumber
\\
&-& 2 e^{-9 \gamma t ({J}-{K})^2}-7\left. \right)+1.
\end{eqnarray}
Note that for the Werner state, the linear entropy is just $
\mathcal{S}_L( \varrho_{\text{iso}}(0))=1-p^2$.
As we are going to consider as the initial states the  whole family of isotropic states, it is illustrative to show first 
(see Fig.\ref{fig:the Qutrit-Qutrit Isotropic}), how entanglement and linear entropy depend on the initial purity $p$ of the state.

Fig.\ref{fig:the Qutrit-Qutrit Isotropic} clearly illustrates that as the mixing parameter $p$ increases, entanglement negativity follows suit, ultimately reaching its zenith at $p=1$. When $p=1$, the isotropic state \eqref{state0} evolves into the maximally entangled state $|\psi\rangle\langle\psi|$. Within the parameter range of $0 < p \leq \frac{1}{4}$, the state remains devoid of entanglement. For a more comprehensive understanding of this region, detailed insights can be found in the work of the authors in \cite{PhysRevA.104.032423}, where they expound upon the quantum nature within the separable region. In the parameter range of $\frac{1}{4} < p \leq 1$, the state manifests non-zero entanglement, featuring a significant point of intersection at $p \approx 0.6672$. Furthermore, when $p$ is set to zero, the isotropic state \eqref{state0} undergoes a transformation into the maximally mixed state $\mathbb{I}_9 / 9$ characterized by a linear entropy of $1$. In summary, a decrease in linear entropy, signifying reduced mixing and disorder, coincides with an increase in entanglement.
\section{Results and discussions}\label{sec4}
Let us start by reviewing the effect of decoherence in the absence of external magnetic fields for a fixed initial purity of the Werner state and fixed values of the Hamiltonian parameters. 
\begin{figure}[ht]
    \centering
    \begin{subfigure}[t]{0.45\textwidth}
        \centering
        \subcaption{} 
        \includegraphics[width=0.9\textwidth]{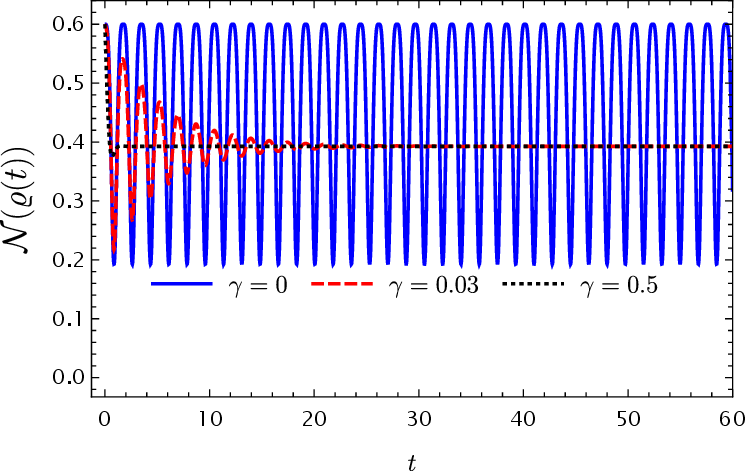}
        \label{Fg8a}
    \end{subfigure} 

     \begin{subfigure}[t]{0.45\textwidth}
        \centering
        \subcaption{} 
        \includegraphics[width=0.9\textwidth]{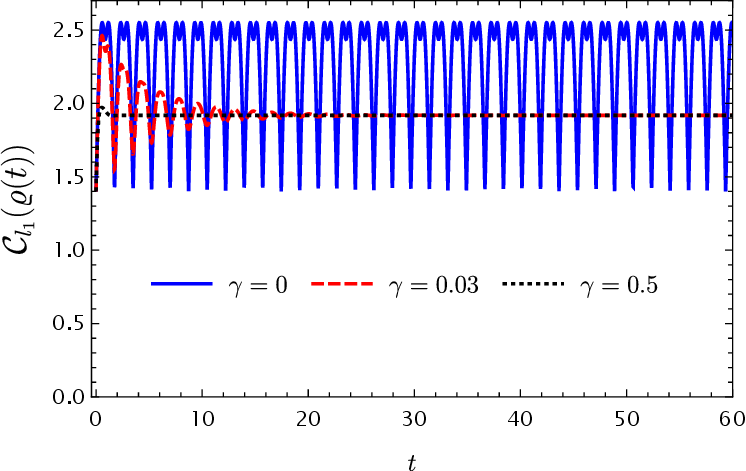}
        \label{Fg8b}
    \end{subfigure}
    
    \begin{subfigure}[t]{0.45\textwidth}
        \centering
        \subcaption{} 
        \includegraphics[width=0.9\textwidth]{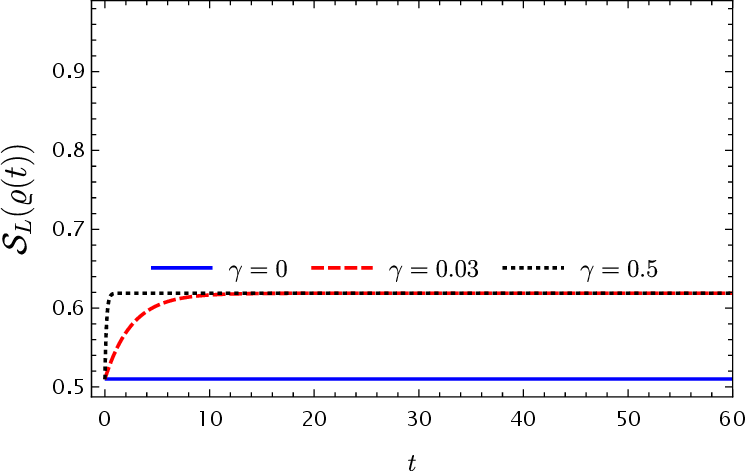}
        \label{Fg8c}
    \end{subfigure}
    
    \caption{\raggedright Dynamics of negativity \ref{Fg8a}, linear entropy \ref{Fg8b}, and $\mathcal{C}_{l_{1}} (\varrho(t))$ \ref{Fg8c}, for fixed values of ${J}=0.8$, ${K}=-0.4$, ${B_z}= 0$, $ p= 0.7$, and versus different intrinsic decoherence rates $\gamma$.}
    \label{Fig.2}
\end{figure}

 Milburn’s decoherence model, with zero decoherence rate ($\gamma = 0$), results in a unitary evolution of the density matrix $\varrho(t)$, which presents periodic oscillations over time. Both negativity and coherence (see Figs.\ref{Fg8a}-\ref{Fg8b}) follow the oscillatory behaviour while the linear entropy (see Fig.\ref{Fg8c}), which depends solely on the initial state, remain invariant in the absence of intrinsic decoherence. Introducing a finite decoherence rate ($\gamma > 0$) induces damping on the quantum dynamics. The system converges to a steady state with decreasing amplitude of oscillations over time. The larger the decoherence rate, the faster the steady state is reached, highlighting the destructive role of intrinsic decoherence. 
 It is important to remark that independently of the value of the decoherence parameter $\gamma$, the steady state reached in all cases coincides, (see Fig.(\ref{Fig.2}), displaying equal values on the negativity, coherence, and linear entropy. Surprisingly enough, the steady state remains quite robust towards decoherence and relatively independent of it.  The reasons for such robustness are related to the large degeneracy of the energy eigenstates of the Hamiltonian in absence of external magnetic fields, as we shall analyze later.
\begin{figure}[ht]
    \centering
    \begin{subfigure}[t]{0.45\textwidth}
        \centering
        \subcaption{} 
        \includegraphics[width=0.9\textwidth]{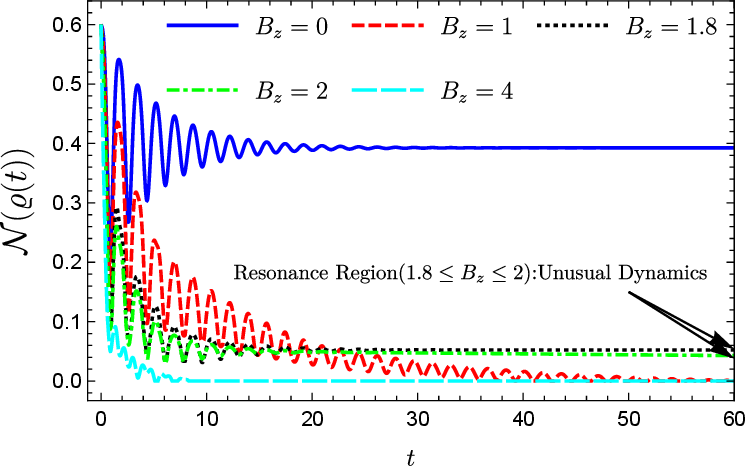}
        \label{B1}
    \end{subfigure} 
    
    \begin{subfigure}[t]{0.45\textwidth}
        \centering
        \subcaption{} 
        \includegraphics[width=0.9\textwidth]{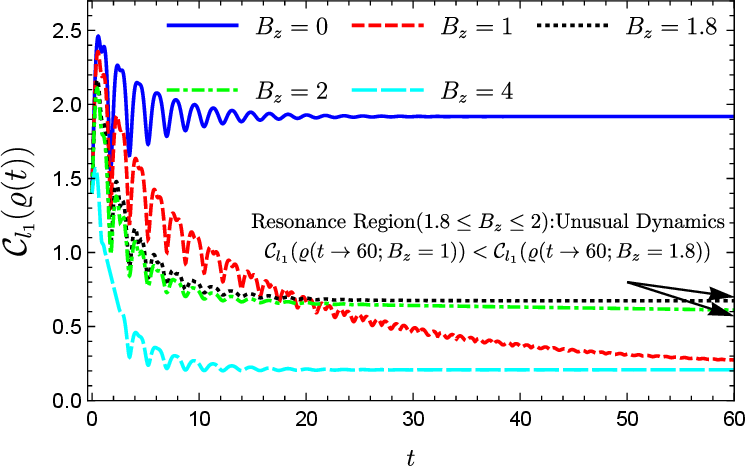}
        \label{B2}
    \end{subfigure}
    
    \begin{subfigure}[t]{0.45\textwidth}
        \centering
        \subcaption{} 
        \includegraphics[width=0.9\textwidth]{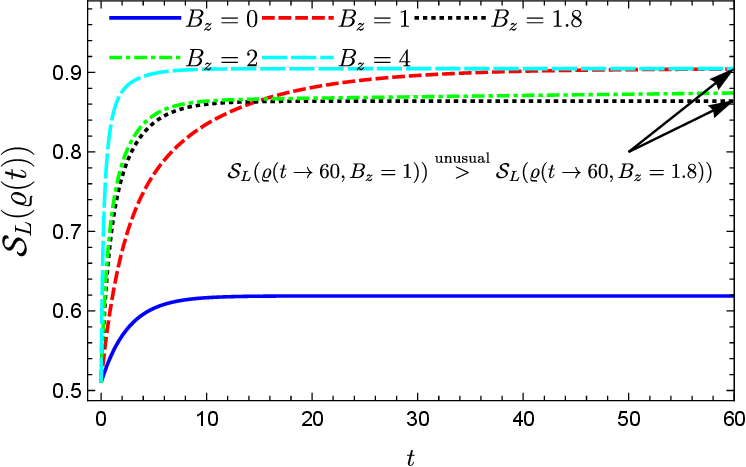}
        \label{B3}
    \end{subfigure}

    \caption{\raggedright Dynamics of negativity \ref{B1}, linear entropy \ref{B2}, and $\mathcal{C}_{l_{1}} (\varrho(t))$ \ref{B3}, for fixed values of ${J}=0.8$, ${K}=-0.4$, $ p= 0.7$, $\gamma=0.03$, and versus different magnetic field strengths ${B_z}$.}
    \label{Fig.3}
\end{figure}

Let us now discuss the effects of an external magnetic field on the system's dynamics for a moderate damping rate of $\gamma=0.03$.  Our results are summarized in Fig. \ref{Fig.3}, displaying the dynamics of negativity Fig. \ref{Fig.3}(a), quantum coherence \ref{Fig.3}(b), and linear entropy (3c). For the easiness of comparison, we have included as well the case $B_z=0$.

As it becomes clear, adding an external magnetic field becomes detrimental to preserve quantum properties. Already for ${B_z}\simeq 1$, the system reaches a steady state where neither entanglement nor quantum coherence are present and, consequently, mixedness stabilizes at $\mathcal{S}_L (\varrho(t\to \infty))\approx 0.9$, very close to the maximally mixed state. The same result is obtained by increase values of the external magnetic field, as shown in Fig.\ref{Fig.3} for ${B_z}=4$. 
 
\begin{figure}[ht]
    \centering{
    \begin{subfigure}[t]{0.45\textwidth}
        \centering
        \subcaption{} 
        \includegraphics[width=0.9\textwidth]{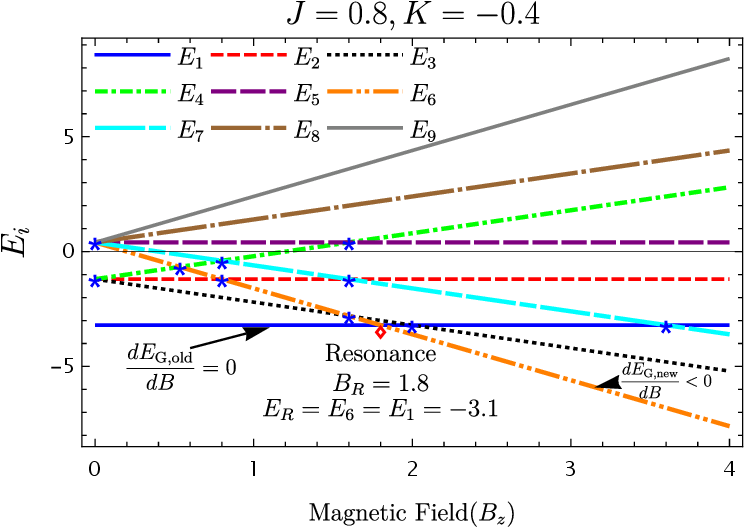}
        \label{E1}
    \end{subfigure} 
    \begin{subfigure}[t]{0.45\textwidth}
        \centering
        \subcaption{} 
        \includegraphics[width=0.9\textwidth]{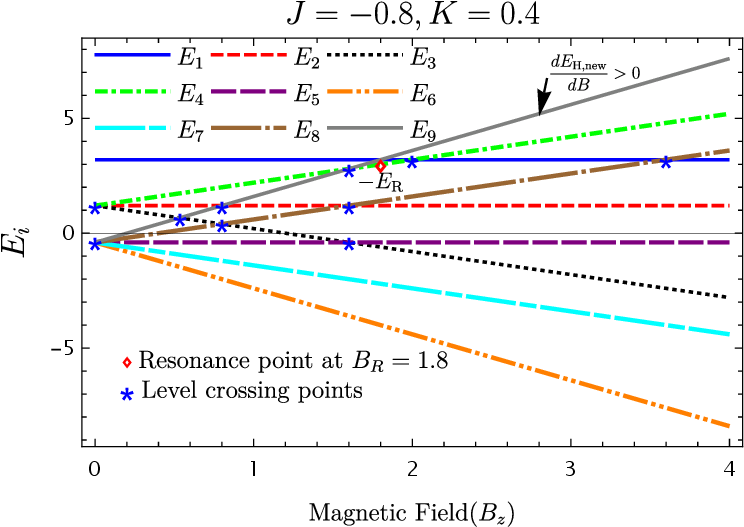}
        \label{E2}
    \end{subfigure}
     
    }
    \caption{\raggedright Energy Levels $E_i$ as a function of $B_z$ with the ground state $(E_{G})$ and the highest energy State $(E_{H})$. $J>0, K<0$ \ref{E1} Crossing Points against $J<0, K>0$ \ref{E2}: Behavior Near Resonance.}
    \label{Fig.6}
\end{figure}

Despite the magnetic field destroys the quantum features of the system, a remarkable abnormal behaviour manifest for an external magnetic field around $B_z=1.8$, where steady-state entanglement and quantum coherence increase and amplify unexpectedly. The same behavior is reflected in the linear entropy, indicating enhanced quantum robustness at this specific magnetic field value. To explore the mechanisms behind the non-trivial behavior, we turn to Fig.\ref{Fig.6}, which illustrates the eigenvalue spectra of the Hamiltonian  
as a function of the external magnetic field. As observed, the Hamiltonian eigenstates are highly degenerate in the absence of an external magnetic field. Such degeneracy highly prevents damping and decoherence, as indicated by Eq. \ref{sol}. As soon as an external magnetic field acts on, the degeneracy is highly lifted and damping and decoherence become very relevant. Thus, energy degeneracies at critical magnetic field values are essential for inducing robustness in front decoherence, enhancing quantum coherence and entanglement.  
Specifically, for the spin-1 model presented here, degeneracy occur at $B_z=\pm\frac{3}{2}(K - J)$(see the appendix \ref{appendix:energy_basis}). Moreover, due to the high symmetry of the Hamiltonian, a resonance is hit at the same value of the magnetic field, both for $J>0$ and  $K<0$ (see Fig.\ref{E1}), or for $J<0$ and  $K>0$ ( Fig.\ref{E2}). In the first case, the resonance occurs for the lowest energy eigenstates, in the later for the largest ones. 

\begin{figure}[ht]
    \centering{
    \begin{subfigure}[t]{0.45\textwidth}
        \centering
        \subcaption{} 
        \includegraphics[width=0.9\textwidth]{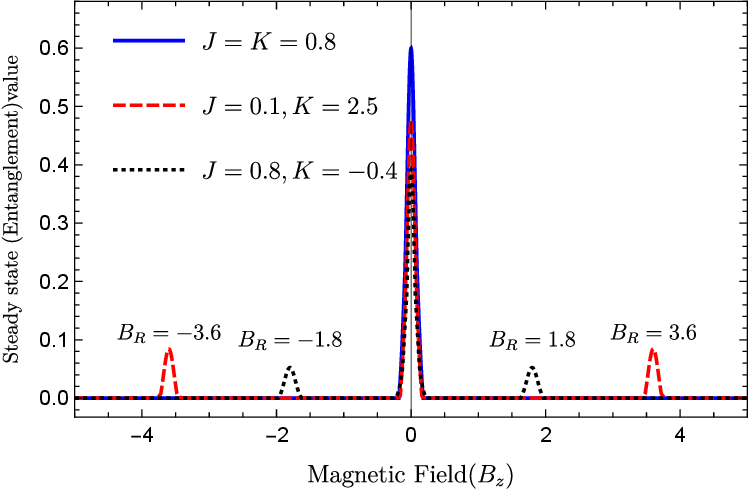}
        \label{Fig4b}
    \end{subfigure}
    
    \begin{subfigure}[t]{0.45\textwidth}
        \centering
        \subcaption{} 
        \includegraphics[width=0.9\textwidth]{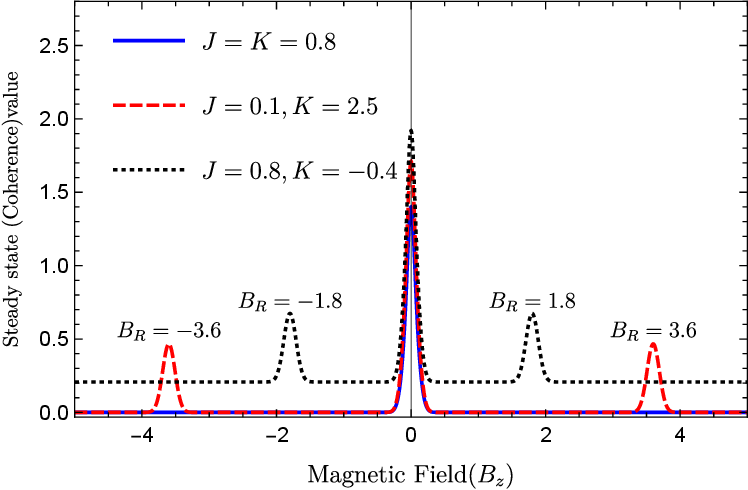}
        \label{Fig4a}
    \end{subfigure} 
    }
    \caption{\raggedright Steady-state entanglement and coherence as functions of ${B_z}$ for fixed values of $ p= 0.7$, $\gamma=0.03$, and anomalies at resonance points.}
    \label{Fig.5}
\end{figure}

To further demonstrate the effect of the resonances induced by external magnetic fields we show in  Fig.\ref{Fig.5}, the entanglement and quantum coherent in the steady state for different values of the Hamiltonian couplings $J$ and $K$. 
Regardless the values of $J$ and $K$, in the absence of an external magnetic field, the Hamiltonian degeneracy leads to preservation of quantum properties of the initial state. The larger the degeneracy is, the stronger the robustness of the state to decoherence effects. Resonances appear always at values $B_z=\pm \frac{3}{2}(K-J)$, as demonstrated there. As can be seen, at all resonance points $B_R$, steady-state coherence, and entanglement persist, reflecting the resilience of quantum properties under resonance conditions. 

\vspace{-0.5cm} 
\section{Concluding remarks}\label{sec5}
 
In most experimental cases, magnetic fields are unavoidable and difficult to shield, enhancing decoherence and damping and leading to detrimental preservation of quantum features, In this work, we have shown, however, that external magnetic fields can also be used to stabilize quantum features. 
Our system of study has been a two spin spin-1 particles in a generalized Heisenberg potential confined into a double-well potential, under the action of intrinsic decoherence model (ID). Despite the limitations of ID models, they provide a very useful tool to analyze the effects of damping and decoherence without considering open quantum systems and serve as a guide for more realistic situations. In such a case, we have shown that control of the 
external magnetic field might lead to resonance between Hamiltonian energy eigenstates. These resonances, effectively enhance the system’s stability, mitigates decoherence, and reduce disorder. Although our results depend on our initial state, which is an isotropic Werner state, the results here presented are general and should equally apply to other Hamiltonians and other initial states under ID models. 
\vspace{-1.76cm} 
\section*{Appendix: Density Matrix in the Energy Eigenbasis}
\label{appendix:energy_basis}
The evolved density matrix in Equation \eqref{sol}, when expressed in the \textbf{energy eigenbasis} of the Hamiltonian \( \{|E_i\rangle\}_{i=1}^9 \), can be shown as a direct sum of a \( 4 \times 4 \) block for the coupled energy levels and individual diagonal terms for uncoupled states and takes the following form.
\[
\resizebox{\columnwidth}{!}{
$\varrho(t)=
\begin{pmatrix}
	\rho_{11}^{(E)} &\rho_{15}^{(E)} & \rho_{16}^{(E)} & \rho_{19}^{(E)} & 0 & 0 & 0 & 0 & 0 \\
	(\rho_{15}^{(E)})^* & \rho_{55}^{(E)} & \rho_{56}^{(E)} & \rho_{59}^{(E)} & 0 & 0 & 0 & 0 & 0 \\
	(\rho_{16}^{(E)})^* & (\rho_{56}^{(E)})^* & \rho_{66}^{(E)} & \rho_{69}^{(E)} & 0 & 0 & 0 & 0 & 0 \\
	(\rho_{19}^{(E)})^* & (\rho_{59}^{(E)})^* & (\rho_{69}^{(E)})^* & \rho_{99}^{(E)} & 0 & 0 & 0 & 0 & 0 \\
	0 & 0 & 0 & 0 & \rho_{22}^{(E)} & 0 & 0 & 0 & 0 \\
	0 & 0 & 0 & 0 & 0 & \rho_{33}^{(E)} & 0 & 0 & 0 \\
	0 & 0 & 0 & 0 & 0 & 0 & \rho_{44}^{(E)} & 0 & 0 \\
	0 & 0 & 0 & 0 & 0 & 0 & 0 & \rho_{77}^{(E)} & 0 \\
	0 & 0 & 0 & 0 & 0 & 0 & 0 & 0 & \rho_{88}^{(E)}
\end{pmatrix}$
}
\]
\begin{align*}=
\varrho_{[1,5,6,9]} \oplus \rho_{22}^{(E)} \oplus \rho_{33}^{(E)} \oplus \rho_{44}^{(E)} \oplus \rho_{77}^{(E)} \oplus \rho_{88}^{(E)},
\end{align*}
The diagonal elements remain constant over time, depending only on the purity $p$ of the initial state, whereas the off-diagonal elements evolve to reflect the coherence dynamics, and this off-diagonal element shows that for $B_z=\pm \frac{3}{2}(K - J)$, the argument of the exponential terms in the matrix elements approaches zero as time goes to infinity, ensuring the preservation of quantum properties and enhancing the robustness of quantum features in the two qutrit system under intrinsic decoherence. Here, the elements of the density matrix are expressed as follows:
\[
\resizebox{\columnwidth}{!}{
$\rho_{11}^{(\text{E})}=\frac{1}{9},  
\rho_{22}^{(\text{E})}=\rho_{33}^{(\text{E})}=\rho_{44}^{(\text{E})}=\rho_{77}^{(\text{E})}=\rho_{88}^{(\text{E})}=\frac{1-p}{9}, \rho_{55}^{(\text{E})}=\frac{p+1}{9}, $
}
\]
\vspace{-0.6cm}
\[
\resizebox{\columnwidth}{!}{
$\rho_{66}^{(\text{E})}=\rho_{99}^{(\text{E})}=\frac{2p+1}{9} ,\rho_{15}^{(\text{E})}=-\frac{p}{9} \sqrt{2}  e^{\frac{3}{2} t (J-K) (3\gamma(K-J)+2 i)},$
}
\]
\vspace{-0.6cm}
\[
\resizebox{\columnwidth}{!}{
$\rho_{16}^{(\text{E})}=-\frac{p \exp \left(-\frac{1}{2} t (2 B_z-3(J-K)) (\gamma(2B_z-3(J-K))+2 i)\right)}{3 \sqrt{3}},$
}
\]
\vspace{-0.6cm}
\[
\resizebox{\columnwidth}{!}{
$\rho_{19}^{(\text{E})}=-\frac{p \texttt{{\cal b}}^*}{3\sqrt{3}},(\rho_{56}^{(\text{E})})^*=\rho_{59}^{(\text{E})},\rho_{69}^{(\text{E})}=\frac{p}{3} e^{4 B_z t (-2B_z\gamma+i)},$
}
\]
\vspace{-0.6cm}
$\rho_{56}^{(\text{E})}= \frac{p}{3}\sqrt{\frac{2}{3}} \texttt{{\cal a}}.$
\bibliography{reference}
\bibliographystyle{apsrev4-2}

\end{document}